\documentclass[%
 reprint,
 amsmath,amssymb,
 aps,
]{revtex4-1}

\usepackage{graphicx}
\usepackage{dcolumn}
\usepackage{bm}
\usepackage{xspace}
\usepackage{lineno}

\usepackage{xcolor}

\newcommand{\pt}{\ensuremath{p_{\rm T}}\xspace}
\newcommand{\ipp}{\ensuremath{I_{\rm pp}}\xspace}
\newcommand{\rt}{\ensuremath{R_{\rm T}}\xspace}

\newcommand{\nmpi}{\ensuremath{N_{\rm mpi}}\xspace}

\newcommand{\py}{PYTHIA\xspace}

\begin{document}

\preprint{APS/123-QED}

\title{Apparent modification of the jet-like yield in proton-proton collisions \\with large underlying event}
\author{Gyula Benc\'edi}
\author{Antonio Ortiz}
 \email{antonio.ortiz@nucleares.unam.mx}
\author{Sushanta Tripathy}%
\affiliation{%
Instituto de Ciencias Nucleares, Universidad Nacional Aut\'onoma de M\'exico,\\
 Apartado Postal 70-543, M\'exico Distrito Federal 04510, M\'exico 
}%

\date{\today}

\begin{abstract}
This paper presents the production of charged particles associated with high-$p_{\rm T}$ trigger particles ($8<p_{\rm T}^{\rm trig.}<15$ GeV/$c$) 
at midrapidity in proton-proton collisions at $\sqrt{s}=5.02$\,TeV simulated with \py 8.244. The study is performed as a function of the relative transverse activity 
classifier, $R_{\rm T}$, which is the relative charged-particle multiplicity in the transverse region ($\pi/3<|\phi^{\rm trig.}-\phi^{\rm assoc.}|<2\pi/3$) 
of the di-hadron correlations, and it is sensitive to the Multi-Parton Interactions. The evolution of the yield of associated particles on both the towards 
and the away regions ($3\leq p_{\rm T}^{\rm assoc.}< 8$\,GeV/$c$) as a function of $R_{\rm T}$ is investigated. We propose a strategy which allows for the modelling and subtraction 
of the underlying event contribution from the towards and the away regions in challenging environments like those characterised by large $R_{\rm T}$. We found that 
the signal in the away region becomes broader with increasing $R_{\rm T}$, while its corresponding yield is independent of $R_{\rm T}$. Contrarily, the yield increases with $R_{\rm T}$ in 
the towards region. This effect is reminiscent of that seen in heavy-ion collisions, where an enhancement of the yield in the towards region for 0-5\% central 
Pb--Pb collisions at $\sqrt{s_{\rm NN}}=2.76$\,TeV was reported. A discussion on the origin of these effects in \py, as well as their implications for the 
interpretation of recent LHC results for pp collisions, is presented.
\end{abstract}

\maketitle


\section{Introduction}

High-energy proton-proton (pp) collisions exhibit heavy ion-like behavior, namely collectivity~\cite{Khachatryan:2010gv,Khachatryan:2016txc,Acharya:2018orn} 
and strangeness enhancement~\cite{ALICE:2017jyt}, raising the question whether a small drop of Quark-Gluon Plasma (QGP) is produced in such collisions~\cite{Nagle:2018nvi}. 
In nucleus-nucleus (AA) collisions, the in-medium parton energy loss, i.e. the modification of the jet due to the interaction of the fast parton within the dense QCD matter, is among 
the key studied effects which indicates the presence of QGP~\cite{Loizides:2016tew}. To date no jet quenching~\cite{Wiedemann:2009sh} is found in small collision systems, 
however a first attempt has already been made recently to unveil it in high-multiplicity pp collisions at $\sqrt{s}=13$\,TeV~\cite{Jacobs:2020ptj}. 
That study consists of measurements of the semi-inclusive acoplanarity distribution of jets recoiling from a high-\pt trigger hadron. A significant broadening of the recoil 
jet acoplanarity distribution is observed in high-multiplicity pp collisions; the observation is characteristic of jet quenching. However, the effect is seen not only in the experimental 
data but in the \py 8.2 model~\cite{Sjostrand:2014zea} (tune Monash 2013~\cite{Skands:2014pea}), which lacks the mechanism of jet quenching. Therefore, studies using \py 8 with jets in high multiplicity pp events are important in order to understand the potential biases which could mimic jet quenching effects. It is also worth mentioning that recent results which use the 
PYTHIA/Angantyr model~\cite{Bierlich:2018xfw}, to extend the PYTHIA model to heavy-ion collisions, have shown that hadronic re-scattering in such collisions can produce suppression 
patterns even without any partonic energy loss~\cite{daSilva:2020cyn}.
 
In this paper, we explore a quantity called \ipp which is motivated by the $I_{\rm AA}$~\cite{PhysRevLett.97.162301,PhysRevLett.104.252301,Aamodt:2011vg} commonly used in heavy-ion 
collisions to study jet quenching effects. $I_{\rm AA}$ is the ratio of jet-like yield from AA to the one from pp collisions, and it provides the interplay between the parton 
production spectrum, the relative importance of quark-quark, gluon-gluon and quark-gluon final states, and energy loss in the medium in heavy-ion collisions. In the towards region, the $I_{\rm AA}$ 
provides information about the fragmenting jet leaving the medium, while in the away region it additionally reflects the probability that the recoiling parton survives the passage through 
the medium. In Ref.~\cite{Aamodt:2011vg}, the ALICE collaboration studied the $I_{\rm AA}$ for central Pb--Pb collisions. They reported a significant suppression ($I_{\rm AA}\approx0.6$) 
in the away region, whereas a moderate enhancement ($I_{\rm AA}\approx1.2$) was observed in the towards region. It is worth mentioning that the CMS collaboration draw qualitatively similar conclusions 
in Ref.~\cite{CMS-PAS-HIN-12-010}. These effects indicate the presence of the medium produced in heavy-ion collisions with which the partons interact along their path.

In this work,  we report an analogous study in pp collisions as a function of  the underlying event (UE) activity. We study the towards and the away region of the di-hadron correlations of 
charged-particles associated with high-\pt trigger particles ($8<p_{\rm T}^{\rm trig.}<15$ GeV/$c$). The study is based on Monte Carlo (MC) event samples generated using ~\py 8.  The UE is 
the component of the collision which does not directly belong to the leading partonic interaction. At LHC energies, most of this activity originates from semi-hard Multi-Parton Interactions (MPI). 
Therefore, event classifiers sensitive to UE are important to understand the MPI dynamics and their impact on observables used in heavy-ion collisions to characterize the properties of the dense QCD 
medium~\cite{Ortiz:2020rwg,Palni:2020shu}. In our study, we use the Relative Transverse Activity Classifier, \rt, which has been recently introduced~\cite{Martin:2016igp,Ortiz:2017jaz} and 
successfully implemented in the analysis of the LHC data~\cite{Acharya:2019nqn,Zaccolo:2019hxt}. \rt uses the conventional definition of the ``transverse region'', which was adopted in the 
Underlying Event analysis originally introduced by the CDF collaboration~\cite{Field:2000dy,Aaltonen:2015aoa} and applied in experiments at RHIC~\cite{Adam:2019xpp} and at  
LHC~\cite{Khachatryan:2015jza,Aaboud:2017fwp,Acharya:2019nqn}. Using the new variable \rt, our study is aimed at understanding the selection biases which could affect the measurement, 
as well as discussing a strategy to minimize such biases in order to have an observable more suitable for jet quenching searches in small systems. 

The paper is organised as follows:  section 2 is devoted to discuss the main aspects of the \py 8.2 model, section 3 describes the event classifier, \rt, and the procedure to extract the jet-like 
signal using the event mixing technique.  Results are presented in section 4, and finally section 5 contains a summary and outlook.

\section{Event generation with \py 8}

\py is a parton-based event generator, which is one of the most widely used Monte Carlo generators for high-energy collider physics with an emphasis on pp collisions.

The main event of a pp collision can be represented via 2-to-2 matrix elements with hard parton scatterings, defined at the leading order, and it is complemented by the leading-logarithm 
approximation of parton showers including Initial-State Radiation (ISR) and Final-State Radiation (FSR). The UE is everything except the two outgoing hard scattered partons, and it receives 
contributions from the beam-remnants, MPI, ISR and FSR. It is an unavoidable background to most of the studied hadron collider observables. The  hadronization in \py  is  treated  using  the Lund string 
fragmentation model~\cite{Andersson:1983ia}. In the Color Reconnection (CR) picture of the model~\cite{Argyropoulos:2014zoa}, the strings between partons can be rearranged so that the total string 
length is reduced; in effect, the total charged-particle multiplicity of the event is also reduced.

Besides the main event, events with other partonic interaction processes between the incoming partons are expected. These are called MPIs~\cite{PhysRevD.36.2019} and are usually soft in nature, 
although the momentum transfer may also reach the hard interaction energy scale. It was shown earlier that MPIs provide a natural way of explaining the increased activity in the UE in a hard 
scattering~\cite{Skands:2014pea}.

The results reported in this work are based on simulations of inelastic processes using \py version 8.244, which has the Monash 2013 tune as its default, and it will be labelled as Monash in the 
following. This tune has been created for a better description of minimum bias and underlying event observables in high-energy collisions. 

\section{Event activity classifier and jet-like signal extraction}

In order to produce results accessible to the main LHC experiments, we have simulated inelastic pp collisions at $\sqrt{s}=5.02$\,TeV and only primary charged particles with transverse momentum $\pt>0.5$\,GeV/$c$ 
and within pseudorapidity $|\eta|<0.8$ are considered in the analysis. Primary charged particles are defined as all final-state particles including decay products except those from weak decays of strange particles; 
this definition is similar to the one used by the ALICE experiment~\cite{ALICE-PUBLIC-2017-005}. The charged-particle yields are obtained in three different regions defined by the relative azimuthal 
angle, $\Delta\phi=\phi^{\rm trig.}-\phi^{\rm assoc.}$, to the direction of the particle with the largest \pt of the event (trigger particle):

\begin{itemize}
\item Towards region: $|\Delta\phi|<\pi/3$,
\item Transverse region: $\pi/3<|\Delta\phi|<2\pi/3$,
\item Away region: $|\Delta\phi|>2\pi/3$~.
\end{itemize}

The towards and the away regions are dominated by string fragments originating from the hardest partonic process of the event, and are expected to be nearly insensitive 
to the softer UE. However, the transverse region is the most sensitive to UE and therefore can be used to build the event activity classifier \rt~\cite{Martin:2016igp}.
Above a certain \pt threshold ($\pt^{\rm trig.}>8-10$\,GeV/$c$) corresponding to the onset of the UE plateau in the transverse region, the mean charged-particle multiplicity 
in the transverse region, $\langle N_{\rm ch}^{\rm trans.} \rangle$, have less dependence on $\pt^{\rm trig.}$~\cite{Acharya:2019nqn}. Therefore, we focus on events having $\pt^{\rm trig.}$ above the onset of the plateau, and we 
classify events with a trigger particle in the range $8\leq\pt^{\rm trig.}<15$\,GeV/$c$ based on their per-event activity in the transverse region with respect to the mean:
\begin{equation}
\rt=\frac{N_{\rm ch}^{\rm trans.}}{\langle N_{\rm ch}^{\rm trans.} \rangle}~.
\end{equation}

The upper limit on $\pt^{\rm trig.}$ is set in accordance with existing experimental results for central and peripheral Pb--Pb collisions at $\sqrt{s_{\rm NN}}=2.76$\,TeV~\cite{Aamodt:2011vg},
which also guarantees a  regime where jet-like correlations dominate over collective-like effects~\cite{Ortiz:2013yxa,Sjostrand:2018xcd}.  The center-of-mass energy presented in our study has been chosen 
to motivate the development of this analysis using the unique pp, p--Pb and Pb--Pb LHC data at $\sqrt{s_{\rm NN}}=5.02$\,TeV; such analysis would aim at unveiling the onset of jet quenching in high-energy hadronic interactions.

Given that we want to get an insight on how the event selection based on the activity in the transverse region biases the towards and the away region, the study is conducted considering the cases where the correlations 
at partonic level (due to gluon radiation or color reconnection) are disabled. On the one hand, the ``hard scattering'' component consists of the outgoing two ``jets'' plus initial and final-state radiation~\cite{Field:2002vt} 
which contribute to UE~\cite{Aaltonen:2015aoa}. On the other hand, the high-\pt trigger selection also biases the sample towards events with a large amount of gluon radiation which is balanced on the away region by several low-\pt 
jet fragments~\cite{Field:1982vz}.  Color reconnection is a mechanism which also produces correlations among final-state 
particles~\cite{Ortiz:2013yxa,Ortiz:2018vgc}. Therefore, we decided to perform the study for three different settings of the Monash tune: a) without ISR and FSR, b) without CR and c) the default parameters of the Monash tune. 

\begin{figure}[t]
\includegraphics[width=0.5\textwidth]{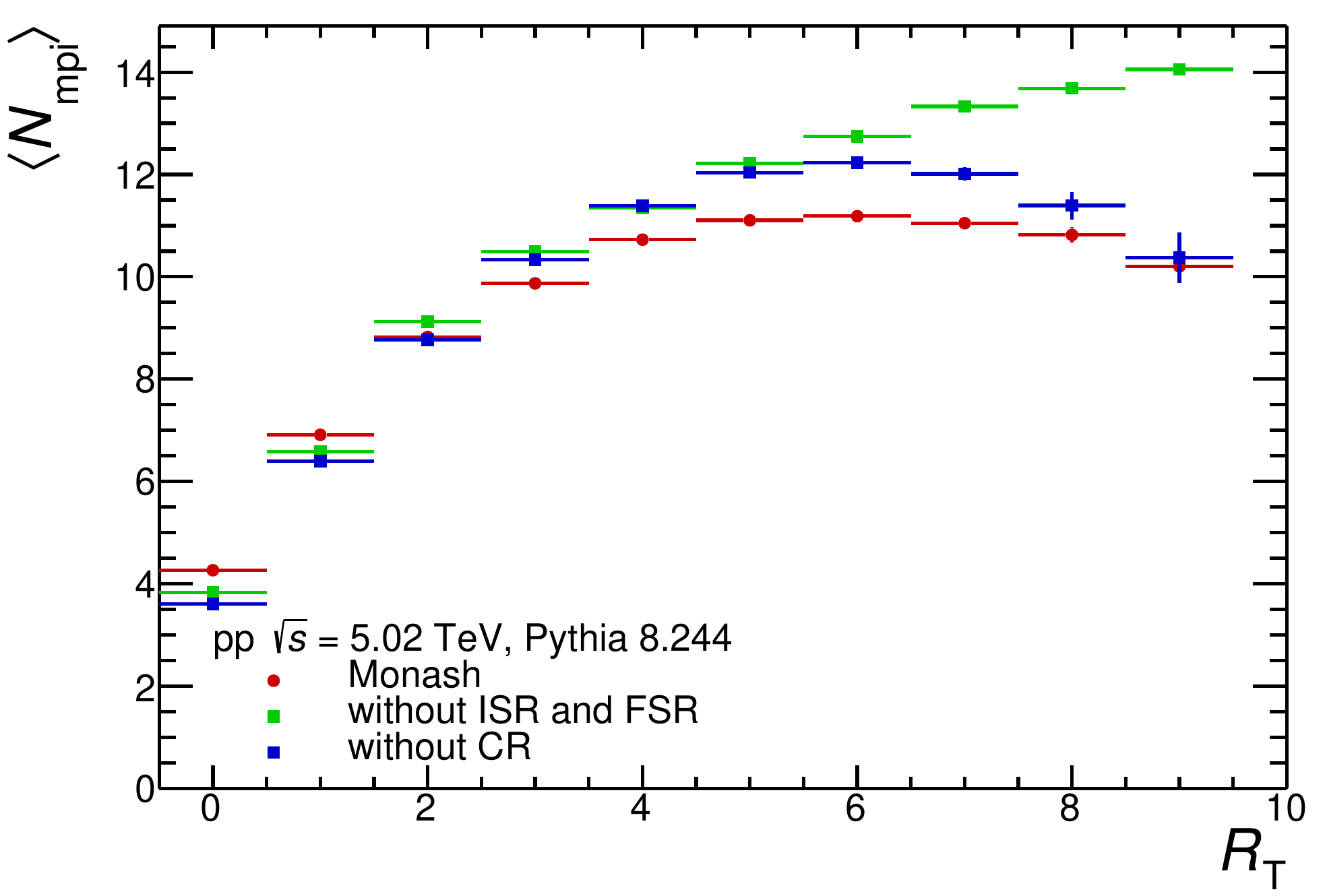}
\caption{\label{fig:1} Average number of Multi-Parton Interactions, $\langle \nmpi \rangle$, as a function of \rt 
in pp collisions at $\sqrt{s}=5.02$\,TeV. The results are shown with default settings, without ISR and FSR, 
and without CR.}
\end{figure}

The correlation between the average number of MPI, $\langle \nmpi \rangle$, and \rt is shown in Fig.~\ref{fig:1}. For $\rt<2$, the average $N_{\rm mpi}$ increases linearly with \rt. It is noteworthy that the correlation 
is nearly independent of ISR, FSR, and CR. Therefore, in this particular \rt region, we can control the amount of MPI and study potential non trivial soft QCD effects on the jet-like region of the di-hadron correlations.
For higher \rt values, $\langle \nmpi \rangle$ exhibits a saturation which is stronger when ISR and FSR are included in the simulations. Having events with the same \nmpi but with different multiplicities in the transverse 
region, suggests that at some point particles from additional jets may be selected to enhance the activity in the transverse region. In turn, above a given \rt value, the activity in the transverse region can be biased 
towards harder processes due to the presence of a third jet in the transverse region~\cite{Field:2002vt}.

\begin{figure*}
\includegraphics[width=0.9\textwidth]{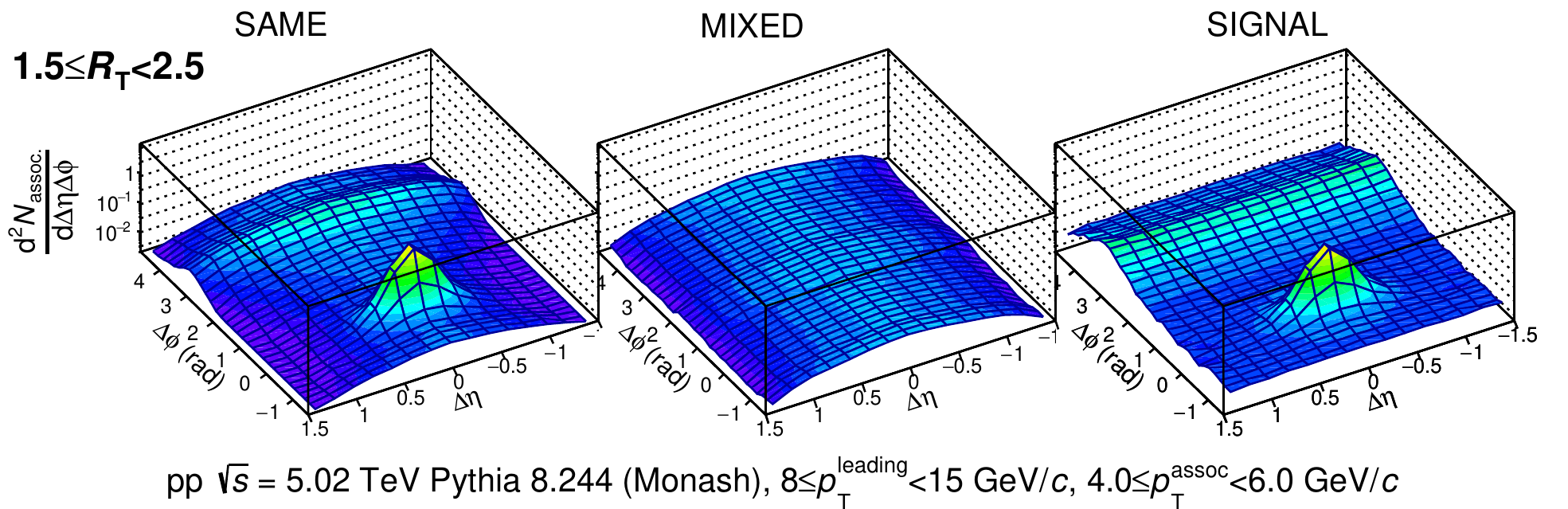}\\
\includegraphics[width=0.9\textwidth]{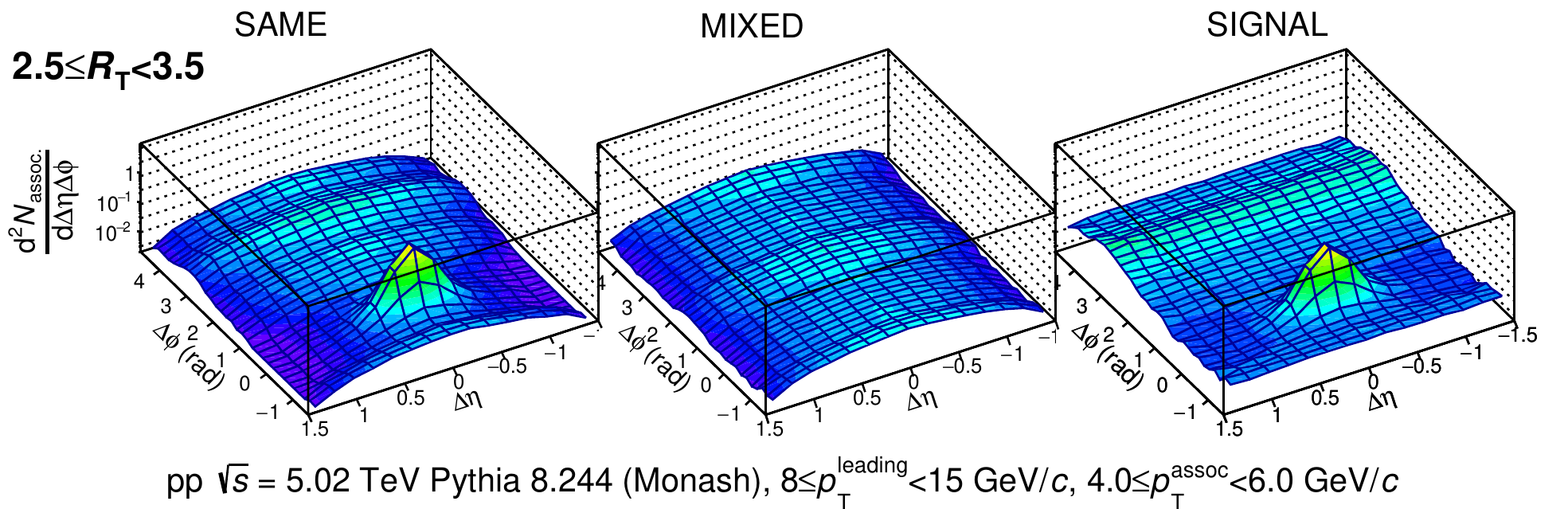}
\caption{\label{fig:2} Charged particle yield as a function of $\Delta\eta$-$\Delta\phi$ for two \rt event classes. 
The associated particles have a \pt in the range $4\leq\pt^{\rm assoc.}<6$\,GeV/$c$. The distribution from the same event (left) exhibits the 
jet peaks in the towards and away regions at $\Delta\phi=0$ and $\Delta\phi=\pi/2$. The structure observed in the transverse region 
can be reproduced using mixed events (middle). The signal (right) is obtained after removing the mixed event distribution 
from the same event distribution using the method defined in Eq.~\ref{eq:2}.}
\end{figure*}

In order to understand the effects of a potential bias, the di-hadron correlation is obtained for the \rt-integrated event class ($\rt\geq0$) as well as for five \rt event classes:   $0\leq\rt<0.5$, $0.5\leq\rt<1.5$, $1.5\leq\rt<2.5$, $2.5\leq\rt<3.5$ and $\rt\geq3.5$ . 
Events  with high \rt ($\geq2.5$) are chosen in order to study the impact on the di-hadron correlations when the MPI saturates (see Fig.~\ref{fig:1}). The left panel of Fig.~\ref{fig:2} shows the di-hadron correlation as a function 
of $\Delta\eta$ ($= \eta^{\rm trigg.}-\eta^{\rm assoc.}$) and $\Delta\phi$. The distribution was built by correlating the trigger particle with associated particles within the same event and with  $4\leq\pt^{\rm assoc.}<6$\,GeV/$c$. 
Correlations for two \rt event classes ($1.5\leq\rt<2.5$ and $2.5\leq\rt<3.5$) are depicted in Fig.~\ref{fig:2}. As expected, the characteristic jet peak at $(\Delta\eta,\Delta\phi)=(0,0)$ and the signal of the associated jet at 
$\Delta\phi=\pi/2$ are observed. In addition, given the selection on \rt, a third structure is present in the transverse region ($\pi/3<|\Delta\phi|<2\pi/3$), and its associated yield increases with \rt. This structure is 
a consequence of the event selection, therefore, its contribution to the towards and the away regions has to be removed. For this purpose mixed events are used to model the uncorrelated contribution as well as the acceptance effect. 
Mixed events are built correlating trigger and associated particles from different events. The mixed events are classified based on the multiplicity of their transverse region. The middle panel of Fig.~\ref{fig:2} shows the 
di-hadron correlation for mixed events. As expected, the activity in the transverse region increases with \rt. By construction, for mixed events, we do not see the jet-like correlations. The jet signal $C(\Delta\eta,\Delta\phi)$ 
is extracted using the following procedure to correct for acceptance effects:

\begin{equation}\label{eq:2}
C(\Delta\eta,\Delta\phi)=B(0,0)\frac{S(\Delta\eta,\Delta\phi)}{B(\Delta\eta,\Delta\phi)}~,
\end{equation} 
where $S$ ($B$) stands for the correlation function for same (mixed) events. The signals after the subtraction of the mixed-events correlation based on the method defined by Eq.~\ref{eq:2} are shown in the right panel of Fig.~\ref{fig:2}. The shape of the towards region 
exhibits little or no dependence on \rt. However, the width of the signal in the away region increases with \rt. The evolution of the observed structures with \rt is further investigated by studying the projection of the $\Delta\eta$-$\Delta\phi$ distribution along the $\Delta\eta$ axis.

Examining the transverse region of the $\Delta\eta$-$\Delta\phi$ distributions one can observe that it is flat in $\Delta\eta$. Therefore, the underlying event contribution to the towards and away regions is subtracted with the zero yield at minimum (ZYAM) assumption~\cite{Ajitanand:2005jj}.    

\section{Results}

\begin{figure*}
\includegraphics[width=0.9\textwidth]{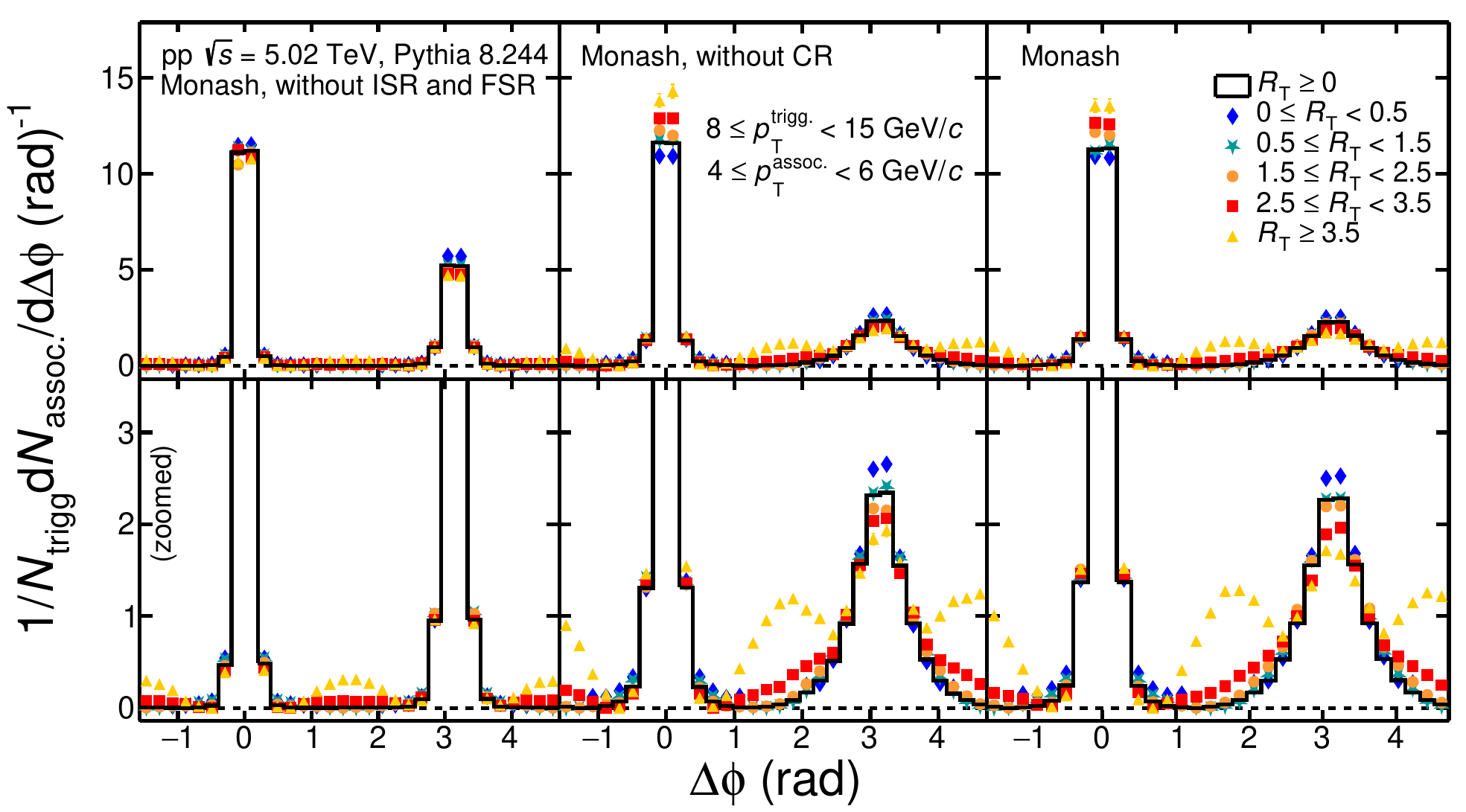}
\caption{\label{fig:3} Charged particle yield as a function of $\Delta\phi$. Results for simulation without radiation (ISR and FSR), without CR, and 
with default settings are shown in the left, middle, and right panels, respectively. \rt-integrated distributions are compared to those from different 
\rt classes. The lower panels are zoomed versions of the upper panels.}
\end{figure*}

The $\Delta\phi$ distributions for various \rt event classes as well as for the \rt-integrated one ($\rt\geq0$) are shown in Fig.~\ref{fig:3}. Although correlations using particles within $|\Delta\eta|<0.8$ and $4\leq\pt^{\rm assoc.}<6$\,GeV/$c$ are shown, the conclusions hold for $3\,{\rm GeV}/c\leq\pt^{\rm assoc.}<\pt^{\rm trig.}$. 
The left panel of Fig.~\ref{fig:3} shows the distributions for simulations without ISR and FSR.  One can observe that the different $\Delta\phi$ regions show no dependence on \rt in the region $0\leq\rt<2.5$. Moreover, it is noteworthy that within statistical uncertainties the shape of the jet peaks are independent of \rt. On the contrary, when ISR and FSR are included in the simulations, the away region exhibits a broadening which becomes more prominent for high \rt values. At the same time, the yield in the towards region increases with \rt. This effect is observed in simulations with (right panel) and without (middle panel) color reconnection. Therefore, CR produces very small effects on the observables which we investigate in the current study. 

Regarding the remaining signal in the transverse region, as anticipated by Fig.~\ref{fig:1}, the presence of a third jet in the transverse region is expected in the $R_{\rm T}$ range where the average MPI activity saturates. In accordance with this observation, for $\rt>2.5$, the distributions exhibit a peak at $\Delta\phi\sim2$\,rad, which is closer to the peak in the away region. The effect is much stronger when ISR and FSR are included in the simulations. We can say that for full simulations the selection on high \rt ($>2.5$) biases the sample towards multi-jet topologies. This effect is observed in experimental data, in particular, for the transverse region, the high-\pt particle production exhibits a strong increase with \rt~\cite{Zaccolo:2019hxt}.

\begin{figure*}[htb]
\includegraphics[width=0.9\textwidth]{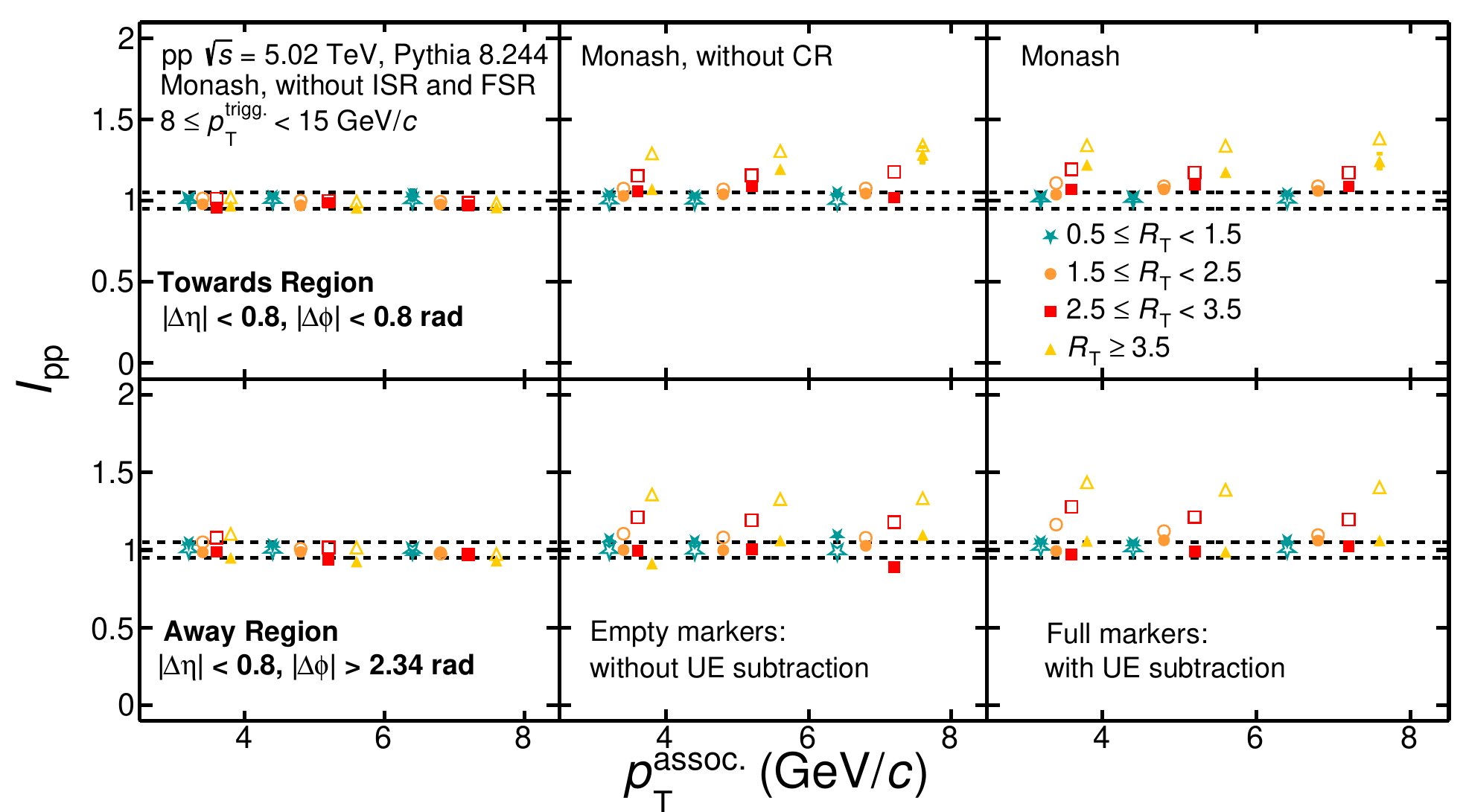}
\caption{\label{fig:4} $I_{\rm pp}$, the ratio of yield from different \rt classes and from the \rt-integrated class as a function of $R_{\rm T}$ for the towards (top) and away region (bottom). 
Results for pp collisions at 5.02\,TeV simulated using the Monash tune are presented for three cases: without ISR and FSR (left), without CR (middle) and the default settings (right). 
Four \rt event classes are shown, as well as three $\pt^{\rm assoc.}$ intervals. \ipp is reported for the cases with (full markers) and without (empty markers) underlying event subtraction.
For better visibility, the data points are slightly displaced on the $\pt^{\rm assoc.}$-axis. Dashed lines in each panel indicate $\pm5\%$ deviation from unity.}
\end{figure*}

In order to quantify the effects discussed above, we calculated the $I_{\rm pp}$ quantity, which is the ratio of yields from different \rt classes to the \rt-integrated one. In the absence of either selection bias or jet quenching, $I_{\rm pp}$  is expected to be consistent with unity. The influence on the jet-like signals of the remaining structures peaked at $\Delta\phi \sim 2$\,rad are reduced by integrating the $\Delta\phi$ distribution within the interval $|\Delta\phi|<0.8$ ($|\Delta\phi-\pi|<0.8$) for the towards (away) region. Figure~\ref{fig:4} shows the $I_{\rm pp}$ as a function of $\pt^{\rm assoc.}$ for both the towards and the away region. Three $\pt^{\rm assoc.}$ intervals are considered: $3\leq\pt^{\rm assoc.}<4$\,GeV/$c$, $4\leq\pt^{\rm assoc.}<6$\,GeV/$c$, and $6\leq\pt^{\rm assoc.}<8$\,GeV/$c$. Results with and without underlying event subtraction are also presented.

\paragraph{Towards region}  In simulations without ISR and FSR, the \ipp is found to be around unity for all the studied $\pt^{\rm assoc.}$ intervals and \rt event classes. Moreover, given the small selection bias discussed before, the subtraction of the underlying event produces a negligible difference. This suggests that the signal extraction method performs well, and it has negligible effect on the results. When ISR and FSR are included in the simulations, the \ipp systematically increases with \rt. The observed rise persists even after the subtraction of the underlying event, although it becomes less steep than in the case without UE subtraction. The evolution of the \ipp with \rt is similar for all the $\pt^{\rm assoc.}$ bins, and it cannot be attributed to the mechanism of color reconnection. Given that this effect is produced by ISR and FSR, and that their contributions are significantly smaller at RHIC than at LHC energies~\cite{Adam:2019xpp}, the enhancement in the towards region is expected to be less pronounced in pp collisions at RHIC energies. This observation is reminiscent of the heavy-ion data, where $I_{\rm AA}$ was found to be around 1.2 for 0-5\% central Pb--Pb collisions at $\sqrt{s_{\rm NN}}=2.76$\,TeV~\cite{Aamodt:2011vg}, and was not observed at RHIC energies~\cite{PhysRevLett.97.162301,PhysRevLett.104.252301}. Our results suggest that radiation in nucleon-nucleon interactions could play a role to explain the behavior of $I_{\rm AA}$  in the towards region from RHIC to LHC energies.

\paragraph{Away region:}  In simulations without ISR and FSR, the away region behaves similarly as the towards region, i.e., the data points sit around unity for the \ipp, and the effect of the underlying event is negligible. However, in simulations with ISR and FSR, the \ipp evolves, both quantitatively and qualitatively, in a different way with respect to the towards region. Only after the removal of UE from the away region, the \ipp is found to be unity within 5\% for all the $\pt^{\rm assoc.}$ and \rt intervals. This suggests that the event selection bias due to leading particle \pt and \rt selection is negligible for this observable. In particular, given the absence of any parton-energy loss mechanism in \py~8, the \ipp is expected to be unity in events with high UE. On the contrary, in central heavy-ion collisions $I_{\rm AA}$ was found to be significantly smaller than one as a consequence of parton energy loss. 

Our proposed strategy allows for the extension of the existing results for pp collisions reported by the ALICE experiment at the LHC~\cite{IppALICE}, which currently covers a very limited $\rt \leq 1.5$ interval. This is because in the quoted experiment, the event classification is done with forward detectors instead of the direct measurement of \rt at midrapidity which causes a limited multiplicity reach~\cite{Acharya:2019mzb}. In the measured \rt interval, \ipp is consistent with unity both in the towards and the away regions, suggesting the absence of jet quenching in small systems. Nevertheless, with our approach, the jet quenching search can be extended up to around $\rt=4$.

\section{Conclusions}
In this work, we have investigated the azimuthal di-hadron correlations as a function of the so-called Relative Transverse Activity Classifier. High-\pt particles associated to trigger particles ($8<\pt^{\rm trig.}<15$\,GeV/$c$) are considered in the study.  We propose to use mixed events followed by a ZYAM subtraction in order to remove the underlying event contribution from the towards and the away regions. The evolution of the jet-like yields with \rt is quantified by the $I_{\rm pp}$ quantity, which is expected to be consistent with unity in the absence of medium effects or selection biases.  We observe that the di-hadron correlations, and therefore \ipp, are independent of \rt in simulations which do not include ISR and FSR. When ISR and FSR are included in the simulations, the yield in the towards region increases with \rt. This effect is reminiscent of the Pb--Pb data from the ALICE and CMS experiments, although in the \py~8 model it is driven by a bias. This bias originates from the interplay of two contributions. On the one hand, the transverse region (\rt) is very sensitive to ISR and FSR, therefore, in events with large \rt (large ISR and FSR) the transverse region often contains a third jet. On the other hand, it has been reported that the high-\pt trigger selection also biases the sample towards events with a large amount of gluon radiation. These effects produce a correlation between the activity in the transverse region and the leading \pt. We emphasize that this selection bias has to be taken into account in any analysis which use \rt as an event classifier. At the same time, the peak in the away region exhibits a broadening which becomes more evident for high \rt values.  The $I_{\rm pp}$ in the away region, on the other hand, is found to be consistent with unity, and independent of \rt and $\pt^{\rm assoc.}$. This suggests that \py~8 with ISR and FSR can partially mimic jet-quenching effects in small systems.

Last but not least, we found that our results are in quantitative agreement with the existing measurements at the LHC, i.e. $\ipp \approx 1$ for $\rt \leq 1.5$, indicating no hint of jet quenching in small collision systems. However, our strategy can pave a way to extend the jet quenching searches up to around $\rt=4$ for small collision systems in experiments at the LHC. 

\begin{acknowledgments}
We acknowledge the technical support of Luciano Diaz and Eduardo Murrieta for the maintenance and operation of the computing farm at ICN-UNAM. Support for this work has been received from CONACyT under the Grant No. A1-S-22917,  PAPIIT-UNAM under Project No.  IN102118. S. T. acknowledges the postdoctoral fellowship of DGAPA UNAM. G. B. acknowledges the postdoctoral fellowship of CONACyT under the Grant No. A1-S-22917, as well as support received during the pandemic by National Research, Development and Innovation Office (NRDIO) OTKA K120660 and FK131979,  2019-2.1.11-TET-2019-00078 (HuMex) and 2019-2.1.6-NEMZKI-2019-00011 (CERN). 
\end{acknowledgments}


\bibliography{biblio}

\end{document}